\documentclass[12pt]{article}
\usepackage[centertags]{amsmath}
\usepackage{amssymb,bm}
\usepackage{graphicx}
\usepackage{subfigure}
\usepackage{psfrag}
\usepackage[mathscr]{eucal}
\def\i{\textrm{i}}
\def\vec#1{\mathbf{#1}}
\renewcommand{\Re}{\textrm{Re}}

\newcommand{\vv}{\mathrm{v}}

\begin{document}

\title{Accuracy of one-dimensional collision integral in the rigid spheres approximation}

\author{O.V. Belai, O.Y. Schwarz, D.A. Shapiro\\
Institute of Automation and Electrometry,\\
Siberian Branch, Russian Academy of Sciences,\\
Novosibirsk 630090, Russia}

\maketitle

\begin{abstract}
The accuracy of calculation of spectral line shapes in
one-dimensional approximation is studied analytically in several
limiting cases for arbitrary collision kernel and numerically in
the rigid spheres model. It is shown that the deviation of the
line profile is maximal in the center of the line in case of large
perturber mass and intermediate values of collision frequency. For
moderate masses of buffer molecules the error of one-dimensional
approximation is found not to exceed 5\%.
\end{abstract}

\section{Introduction}

Modern high resolution spectrometry of molecular gases requires a
precise knowledge of spectral lines shapes.  The present-day
experimental techniques have reached so high precision and
accuracy, that models describing line shapes have to take into
account such fine effects as velocity dependence of collisional
width and shift, correlation between velocity changing and phase
shifting collisions, finite impact time, and radiation relaxation.
The profile of an isolated spectral line is usually obtained by
solving quantum kinetic equation for the off-diagonal element of
the density matrix describing the active gas \cite{ufnRS67}. The
essence of any model describing line shapes is the way it accounts
for the collisions of radiator and perturber molecules. In
principle, in the impact approximation the term accounting for
collisions can be obtained by averaging the corresponding
transition frequencies over velocities of the buffer molecules, as
described in \cite{RS91}. The kinetic equation is a
three-dimensional integral equation. Actually, it can be reduced
to a two-dimensional, because the problem has an axial symmetry
with respect to the wave propagation direction. Due to
computational difficulties, this ab initio approach is rarely used
for approximating experimental line shapes. Instead, most works on
this subject utilize various simplified models of collision term.
The simplest of these models lead to one-dimensional equations.
They are the strong collisions \cite{prNG64}, the weak collisions
\cite{prG61} and the Keilson-Storer \cite{amKS52} models. The more
complicated ones, such as the rigid spheres \cite{KRS72} and
``kangaroo'' \cite{jmpBrissaudFrisch74} models yield 3D equations.

The work \cite{KRS72}, in which the first attempt was made to
proceed from the simplest models to a more complicated one
introduced a combined approach. The rigid spheres collision
kernel, which was analyzed in this work, leads to a 3D integral
equation. However, in order to simplify the numerical
calculations, the kinetic equation was reduced to a 1D integral
equation by averaging the kernel over the transverse components of
velocity with Maxwellian weight. This reduction is equivalent to
adopting the assumption that the distribution of the off-diagonal
element of the density matrix over the transverse (to the wave
propagation direction) velocity components is Maxwellian. In other
words, this approximation, known as the one-dimensional model,
neglects the transfer of nonequilibrium created by the interaction
with the light wave to the distribution over the transverse
velocities. Later the 1D approximation was used in \cite{RS91} and
\cite{praBerman82} to analyze general aspects of influence of
collisions on spectral line profiles. The 1D approximation in the
rigid spheres model was utilized in \cite{praLBB80} to fit the
experimental line shapes. The line shapes in 1D rigid spheres were
comprehensively studied and compared to other models in
\cite{praSM01}.

In our time the calculation of a line shape in the rigid spheres
model can be performed on a usual desktop computer without
utilizing 1D approximation. However, it may still become useful
when improvement of precision of experimental measuring of
spectral lines profiles will persuade researchers to turn to even
more realistic models, with kernels describing simultaneously
dephasing and velocity-changing collisions.

The investigation of precision of the 1D approximation is also
motivated by the following. It has been pointed out recently
\cite{slsWCVT06} that probably much of the disagreement between
theoretical and experimentally measured line shapes could be
removed if the calculation of dephasing term was performed using
the correct velocity distribution of the off-diagonal element of
the density matrix instead of Maxwell's distribution. The
understanding of to what extent this distribution really differs
from equilibrium is required to clarify this problem. The answer
to the question, how much of the non-equilibrium is transported to
the distribution over transverse velocities, could contribute to
such understanding.

The question of precision of the 1D approximation was first
studied quantitatively in \cite{praPS99}. In this article the
problem was studied in the so-called "kangaroo" model. The
accuracy of the 1D approximation was found to be good. Later the
same result was extended to the accuracy of 1D approximation in
describing the light induced drift effect
\cite{jetpParkhomenkoShalagin2000} in the ``kangaroo'' model. This
research was continued in \cite{jetpParkhomenkoShalagin2001} with
special attention to Dicke narrowing effect and gave the same
result. The overall conclusion of these three articles regarding
1D approximation is, its accuracy is high and it can be used for
studying a wide range of problems both in linear and nonlinear
spectroscopy. This thorough analysis does not seem quite general
and comprehensive. The mere fact that the three mentioned simple
models of collision integral lead to 1D kinetic equation (and so
the 1D approximation is precise for them) shows that the accuracy
of this approximation is determined by some fine properties of the
collision integral. Thus the model of collision integral which is
used to study this problem should be more realistic than the
degenerate (in the sense that it has infinite number of
eigenvectors with zero eigenvalue) collision kernel of the
``kangaroo'' model. It seems more appropriate to utilize the rigid
spheres model for this purpose. This model's collision kernel is
obtained by direct averaging of the cross-section (though for a
not very realistic potential), so it has many realistic features.
Another factor that draws attention to this model is that it is
often used (in combination with other terms accounting for
collisional broadening and shifting) to fit experimentally
obtained profiles, e.g. in \cite{praWVCTDM02,praLHCS06}.

The goal of the present work is to study the
precision and the area of applicability of the 1D approximation
quantitatively in
the rigid spheres model. This paper is organized as follows. In
the next section the rigid spheres model is briefly described and
the 1D approximation is introduced. Section \ref{limits} is
devoted to analysis of the accuracy of 1D model in several
limiting cases. The results of numerical calculations in the rigid
spheres model are presented and discussed in
Section~\ref{results}. The conclusions are drawn in the last
section.

\section{Basic equations}

\subsection{The rigid spheres model}\label{billiard}

Let us briefly remind the basic equations defining spectral line
shapes, following \cite{RS91}. The gas interacting with radiation
can be described in terms of density matrix $\hat{\rho}(\vec v)$,
depending on velocity $\vec v$. Let the wave's frequency be close
to resonance with the transition between the levels $m$ and $n$.
Then, in the resonance approximation, the line shape is defined by
the off-diagonal element $\rho_{mn}(\vec v)$. Below we will denote
it as $\rho(v)$ without indices. It is governed by the following
master equation:
\begin{equation}\label{master_omega}
\left(-\i \omega + \i \vec{kv} + \Gamma\right) \rho(\vec v,
\omega) = W(\vec v) + \mathbf{\hat{S}} \rho(\vec v, \omega),
\end{equation}
here $\omega$ is the detuning of the wave's frequency from the
resonance, $\vec k$ is the wave vector of the light wave, $W(\vec
v)$ is Maxwell's distribution, $\Gamma$ is the relaxation
constant, $\mathbf{\hat{S}}$ is the collision operator. The line
shape is given by the formula
\begin{equation}\label{I(omega)}
I(\omega)=\frac{1}{\pi} \;\Re \int  \rho(\vec v) \;d^3 \vec v.
\end{equation}

The relaxation constant $\Gamma$ is assumed to be
velocity-independent. In this case, it influences the line shape
in the following way:
\begin{equation}\label{Gamma}
I(\omega,\Gamma) = \int I(\omega',0) \;\;
\frac{\Gamma}{\Gamma^2+(\omega-\omega')^2} \;
\frac{d\omega'}{\pi},
\end{equation}
that is, the line shape is a convolution of the form of the line
in the absence of $\Gamma$ with a Lorentzian profile. Thus the
relaxation tends to conceal any details of the collision integral,
including the differences between the initial 3D collision
integral and its 1D analog. In this paper we consider equation
(\ref{master_omega}) only with $\Gamma=0$, because in this case
the inaccuracy of the 1D approximation is most noticeable.

We consider the collision operator $\mathbf{\hat{S}}$ in the
impact approximation and also assume that the cross section of
scattering of active molecule by buffer gas molecules is
independent of the molecule's state (full phase memory). Under
these assumptions the collision integral takes the form
\begin{equation}\label{collisions}
\mathbf{\hat S} \rho (\vec v) = \int A(\vec v,\vec v') \rho(\vec
v') d \vec v' - \nu(\vec v) \rho(\vec v).
\end{equation}
The function $A$ is the collision kernel, $\nu$ is the
scattering-out frequency. Owing to the assumption we made, the
functions $\nu$ and $A$ are bound by the following relation:
\begin{equation}\label{nu_via_A}
\nu(\vec v) = \int A(\vec v',\vec v)\; d \vec v'.
\end{equation}
It can be inferred from equations (\ref{collisions}) and
(\ref{nu_via_A}) that
\begin{equation}\label{killIntegral}
\int d \vec v \; \mathbf{\hat S} \;  \rho = 0
\end{equation}
for any $\rho(\vec v)$. Another general property of the collision
integral (\ref{collisions}) can be derived from the detailed
balancing principle:
\begin{equation}\label{detail_balance}
A(\vec v, \vec v') W(\vec v') = A(\vec v',\vec v) W(\vec v).
\end{equation}
Taking into account the definition (\ref{nu_via_A}) of $\nu$  we
obtain
\begin{equation}\label{killMaxwell}
\mathbf{\hat S} W(\vec v) = 0.
\end{equation}

Below we utilize the rigid spheres (or "billiard balls") collision
kernel in all the calculations requiring a definite collision
integral. In this model the differential cross section of
scattering is considered to be independent of relative velocity
and equal to $d\sigma / d o = a^2/4$, here $a$ is the effective
sum of radii of active and buffer molecules, and $d o$ is the
element of solid angle. The collision kernel in the rigid spheres
model is obtained by averaging the probability of a collision,
changing the velocity of active molecule from $\vec v$ to $\vec
v'$, over the velocities of buffer molecules. Its explicit form is
\cite{RS91, KRS72}
\begin{equation}\label{Abilliard} A_{
\mbox{\tiny{\bf RS}}}(\vec v|\vec {v'}) = \frac{N_b\, v_{b
T}}{\sqrt{\pi}} \;\frac{a^2}{ \Delta^2
\zeta}\;\exp\left[-\left(\frac{\boldsymbol{\zeta}(\boldsymbol{\zeta}+\frac{2
\mu}{M}\vec {{v'}})}{\zeta \Delta} \right)^2\; \right],
\end{equation}
here $m$ is the mass of the buffer molecule, $M$ is the mass of
the active molecule, $\boldsymbol{\zeta}=\vec v-\vec {v'}$, $\Delta=\frac{2
\mu}{M} \vv_{bT}$, $\vv_{bT} = \vv_T\left/\sqrt{\beta}\right.$ is
the most probable velocity of the buffer molecule, $\beta = m/M$
is the mass ratio, $\mu={m M}\!\left/{\left(m + M\right)}\right.$
is the reduced mass. The kernel (\ref{Abilliard}) has a
singularity $1/\zeta$ caused by the energy conservation. The
absence of such singularity in most phenomenological kernels
corresponds to suppression of small angle scattering. The kernel
(\ref{Abilliard}) explicitly depends on $\beta$ and demonstrates
correct behavior in the limits $\beta\rightarrow 0$ and
$\beta\rightarrow\infty$.

The scattering-out term in this model has the form:
\begin{eqnarray*}\label{nu}
\nu_{\mbox{\tiny{\bf RS}}}(\vv) = \frac{N_b v_{bT} a^2}{2} \left(
\frac{2}{\sqrt{\pi}}\; e^{- z^2 } + \left[2 z+\frac{1}{z} \right ]
\mbox{Erf} (z) \right),
\end{eqnarray*}
where $\mbox{Erf}(z) = \frac{2}{\sqrt{\pi}}\int\limits^z_0 e^{-x^2} \,dx$
denotes the error function \cite{Bateman} and
$z=\vv\left/\vv_{bT}\right.$.

\subsection{The one-dimensional model}

Equation (\ref{master_omega}) is a two-dimensional (due to axial
symmetry of the problem) integral equation. Its direct numerical
solution presents certain difficulties, so various simplified
models are widely used. One of such models is so-called
one-dimensional approximation. In this models the dependence of
$\rho(\vec v)$ on the transverse components of the velocity $\vec
v$ is considered to be Maxwellian. This assumption allows to write
instead of (\ref{master_omega}) an equation for dependance of
$\rho$ on the longitudinal velocity $\vv_z$:
\begin{multline}\label{master1d}
    \left(-i\omega + i  k  \vv_z +
    \Gamma\right)\rho(\vv_z) = \\ = -
    \nu^{1D} \rho(\vv_z) + \int A^{1D}(\vv_z|\vv'_z) \rho(\vv'_z) d
    \vv'_z,
\end{multline}
where
\begin{equation}\label{A1d}
A^{1D}( \vv_z,\vv'_z)=\int A(\vec {v}|\vec {v'}) \frac{ e^{ - \vec
v'^2_{ \perp }/  \vv_T^2 }}{\pi} d^2 \vec v_{\perp}d^2 \vec
v'_{\perp},
\end{equation}
and
\begin{equation}\label{nu1d}
\nu^{1D}(\vv_z) = \int A^{1D}(\vv'_z, \vv_z) \; d \vv'_z.
\end{equation} Here $A^{1D}$ and $\nu^{1D}$ are one-dimensional collision kernel
and scattering-ou
t frequency, $\vec v_{\perp}$ and $\vec
v'_{\perp}$ are the components of velocity, orthogonal to the wave
propagation direction, and $\vv_T$ is the most probable velocity
of active molecules $\vv_T=\sqrt{2 m T}$.

\begin{figure*}\centerline{
\subfigure[]
{\includegraphics[width=0.3\textwidth]{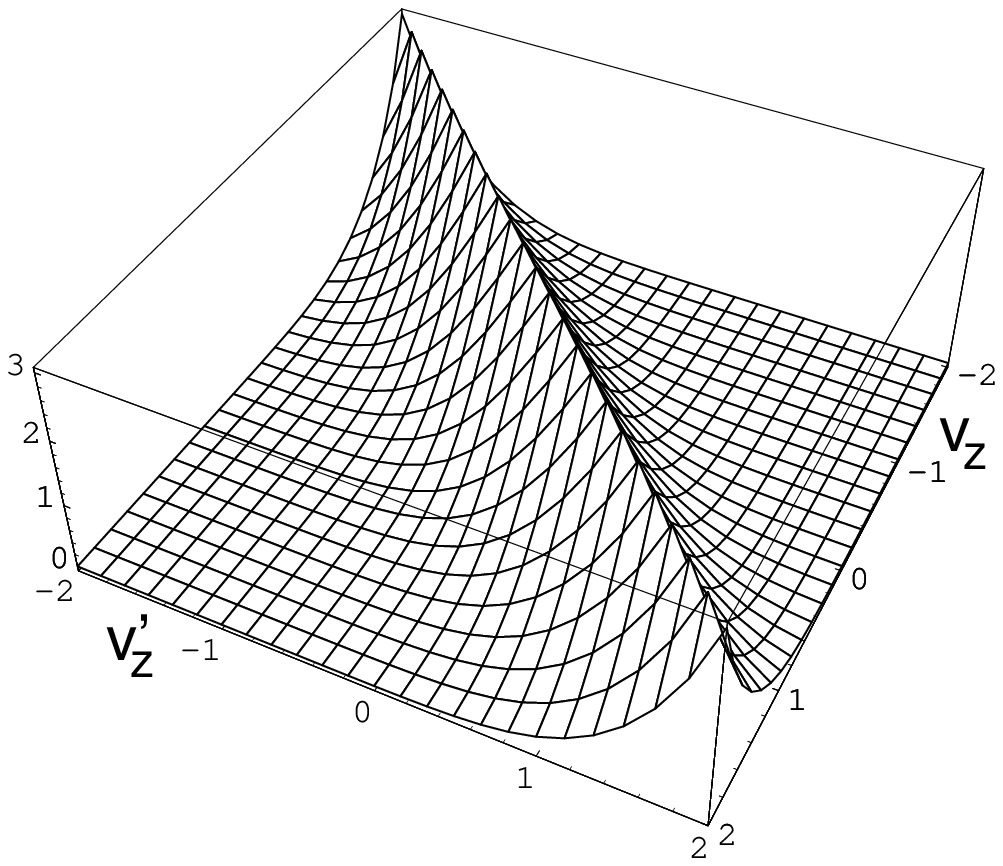}\label{subfig:02}}\hfil
\subfigure[]
{\includegraphics[width=0.3\textwidth]{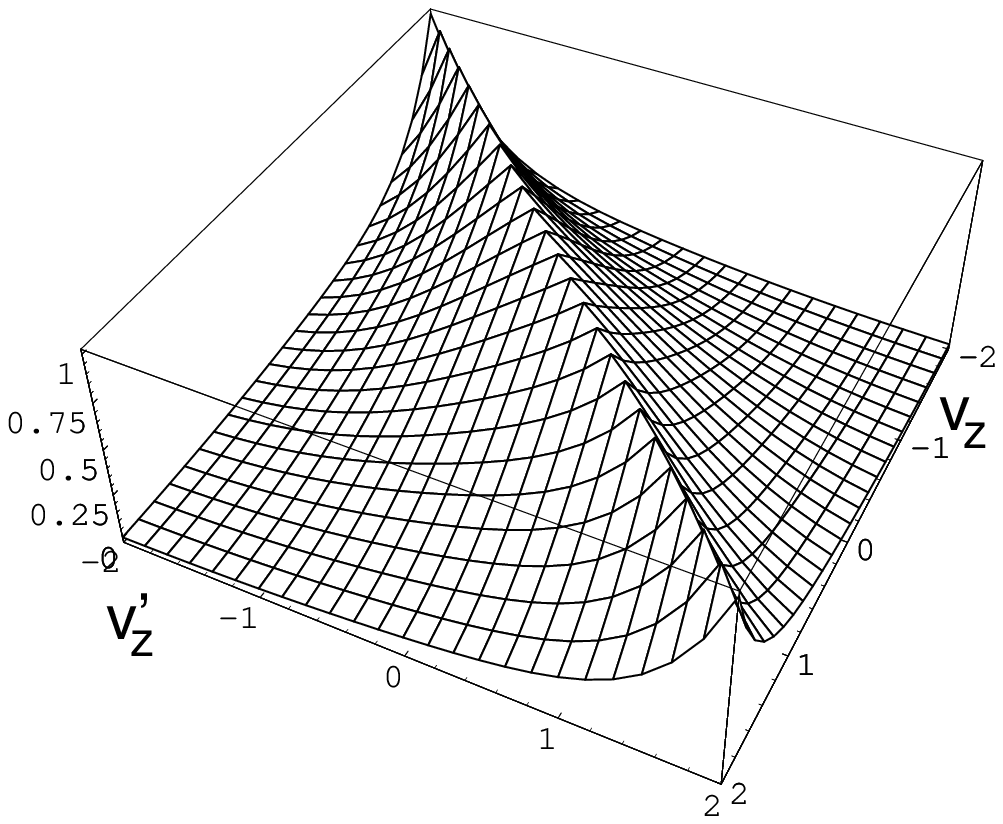}\label{subfig:1}}\hfil
\subfigure[]
{\includegraphics[width=0.3\textwidth]{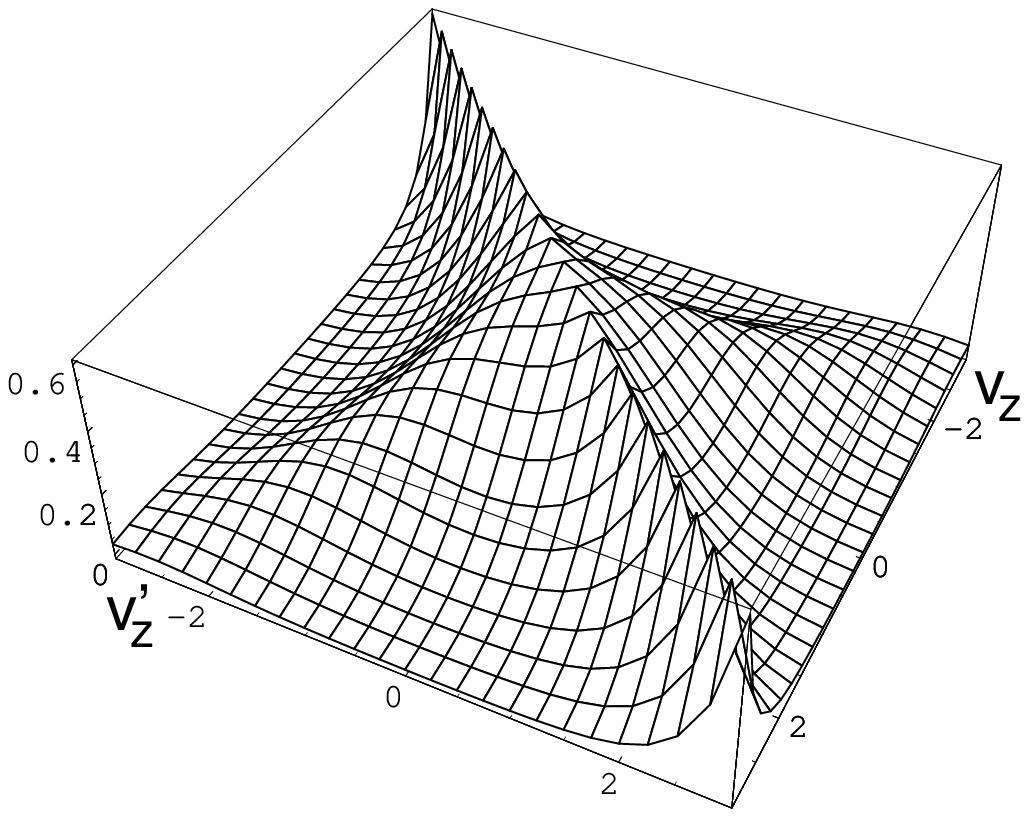}\label{subfig:5}}
}\caption{Shape of the ``symmetrized'' kernel
$A^{1D}(\vv_z,\vv'_z) \exp \left[\left(\vv_z^2 - \vv_z'^2\right)/2
\right]$ at different $\beta=1/5 (a), 1 (b), 5 (c)$.}
\label{fig:kernel}
\end{figure*}

In the 1D approximation the rigid spheres collision kernel
(\ref{Abilliard}) is reduced to
\begin{multline}\label{Abilliard1D}
A^{1D}_{\mbox{\tiny{\bf RS}}}(\vv_z,\vv_z') = \\ = \pi N_b a^2
\vv_T \frac{\beta+1}{4 \beta} \left\{e^{\vv_z'^2-\vv_z^2} \left[
1+\sigma\mbox{Erf}\left(\frac{ \beta-1}{2\sqrt{\beta}} \vv_z
+\frac{\beta+1} {2\sqrt{\beta}} \vv_z'\right)\right] +\right. \\
\left.+1 -\sigma \mbox{Erf}\left(\frac{ \beta + 1}{2 \sqrt{
\beta}} \vv_z + \frac{\beta-1} {2\sqrt{\beta}} \vv_z'\right)
\right\},
\end{multline}
where  $\sigma=\mbox{sign}(\vv_z-\vv_z')$. This kernel as a
function of the initial and final velocity is presented in
fig.~\ref{fig:kernel}. In order to make this figure more
illustrative we plot ``symmetrized'' kernel $A^{1D}(\vv_z,\vv'_z)
exp \left[\left(\vv_z^2 - \vv_z'^2\right)/2 \right]$. This
function is a symmetric function with respect to transformation
$\vv_z\leftrightarrow\vv'_z$. At small perturber mass $\beta\ll1$,
Fig.~\ref{subfig:02}, the kernel function has sharp peak near
$\vv_z'=\vv_z$, which means that the small velocity change is the
most probable. For comparable perturber $\beta\sim1$,
 the peak broadens. For heavy perturber
$\beta\gg1$, Fig.~\ref{subfig:5}, the additional ridge near
$\vv_z'=-\vv_z$ arises corresponding to elastic backward scattering
on a perturbing molecule.

\begin{figure}
\includegraphics[width=0.7\columnwidth]{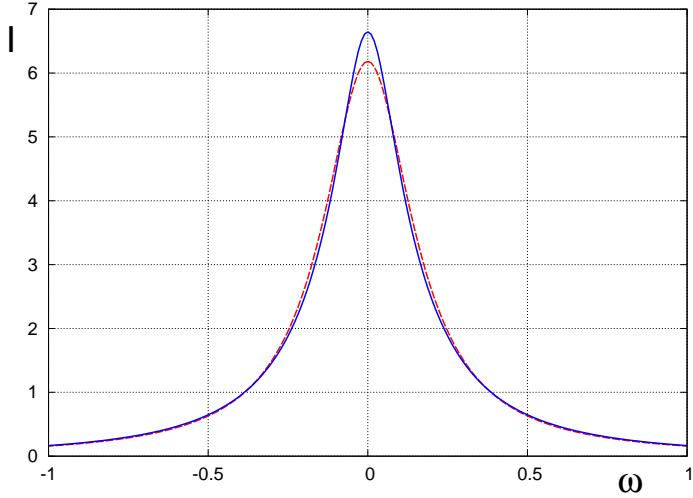}\\
\caption{Shape of the spectral line $I(\omega)$ in the rigid
spheres model at $\beta=100$, $\nu_d=3 k\vv_T$. The solid line
corresponds to the rigid spheres approximation, the dashed line
corresponds to its 1D approximation.}\label{fig:3d1d}
\end{figure}

Integrating expression (\ref{Abilliard1D}) over $\vv_z$, we obtain
the out-scattering frequency of 1D rigid spheres model:
\begin{equation}\label{nu1D}
\nu^{1D}_{\mbox{\tiny{\bf RS}}}(\vv_z) = \pi N_b a^2 \vv_T
\left(\vv_z\; \mbox{Erf} ( \sqrt{ \beta } \vv_z)+\frac{e^{-\beta
\vv_z^2 } }{ \sqrt{ \pi \beta }}+ \frac{1+\beta}{\beta}\;
e^{\vv_z^2} \int_{\vv_z}^{ \infty } \mbox{Erf} (\sqrt{\beta}t) \;
e^{-t^2}\;dt\right).
\end{equation}
The line shape in the 1D rigid spheres model can differ
significantly from its shape in full rigid spheres model, as it is
demonstrated in Fig.~\ref{fig:3d1d}. It can be seen on this
picture, that the 1D line almost coincides with 3D in the wings of
the line, but goes considerably lower in its center. In the next
section we will analyze this difference in several limiting cases.

\begin{figure*}\centerline{
\subfigure[]{\psfrag{long}{\Large $v_z$}\psfrag{trans}{\Large
$v_\bot$}\includegraphics[width=0.45\textwidth]{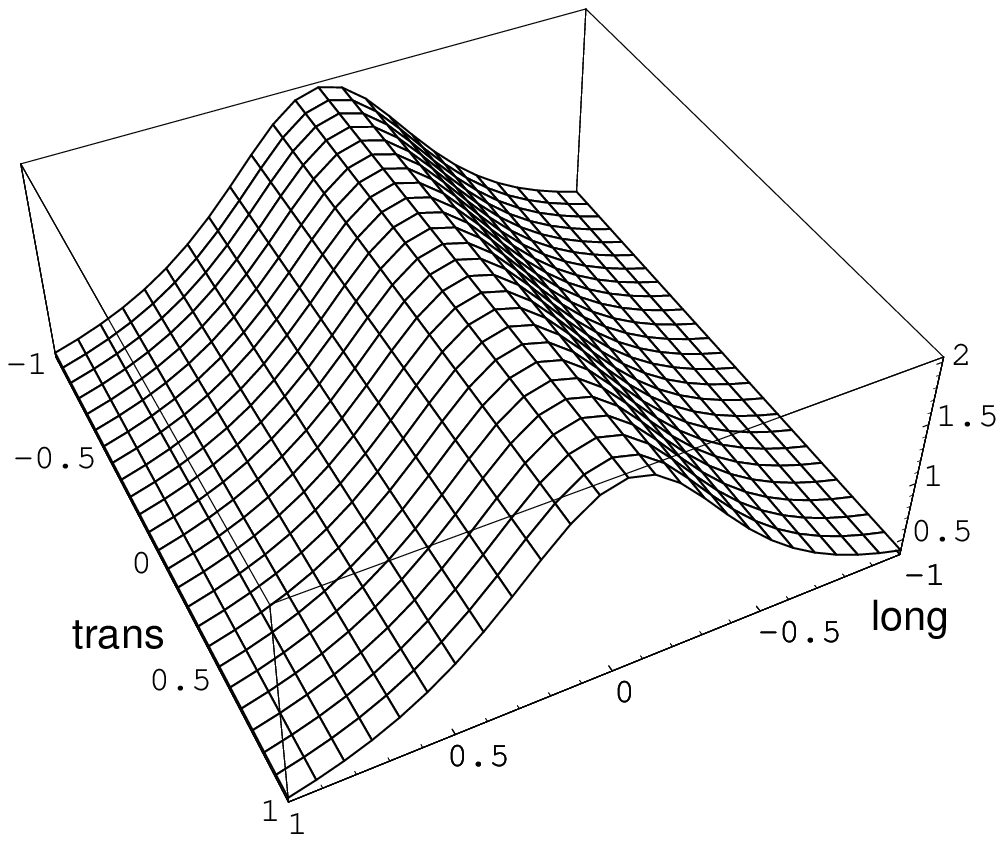}\label{subfig:noransport}}
\subfigure[]{\psfrag{long}{\Large $v_z$}\psfrag{trans}{\Large
$v_\bot$}\includegraphics[width=0.45\textwidth]{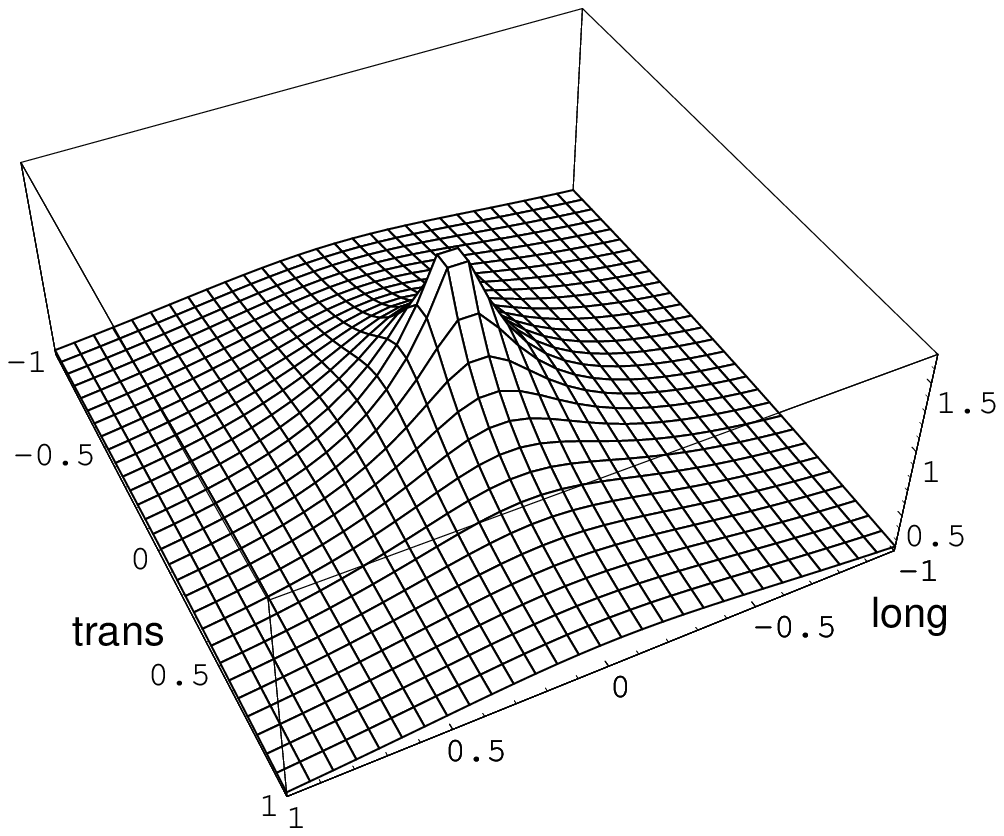}\label{subfig:transport}}
} \caption{Distribution of the value $\rho(\vec v)/W(\vec v)$ over
longitudinal and transverse components of velocity in the rigid
spheres model at $\omega = 0$ and $\Gamma=0.5 \, k \vv_T$. The
first plot (a) corresponds to absence of collisions. The second
plot (b) corresponds to the case $\beta\rightarrow \infty$, $C=3$
(see section \ref{Lorenz}).}\label{fig:ill}
\end{figure*}

The essence of the 1D model is the assumption that the transfer of
non-equilibrium to the distribution of $\rho$ over the transverse
components of velocity is negligible. Figure~\ref{fig:ill}
demonstrates, that this transfer is not weak in general case.
Fig.~\ref{fig:ill} (a) represents the distribution of $\rho(\vec
v)$, divided by Maxwellian distribution, in absence of collisions.
It can be seen that the dependence on transverse velocity is
uniform. Fig.~\ref{fig:ill} (b) represents the distribution of the
same value in presence of the collision integral. The dependence
on the transverse velocity becomes strongly nonuniform. Thus it
could be expected that the impact of transfer of non-equilibrium
on line shapes would be also strong. However, as it will be shown,
in most cases it turns out to be numerically small.

\section{The limiting cases}\label{limits}

\subsection{The wings of spectral line profile}

Let us at first consider the spectral wings of the line. Note that
the calculations below take into account only rigid spheres
elastic collisions and ignore other effects determining the far
wings. It is convenient to turn to the time domain for this
purpose and to introduce the Fourier transform of $\rho(\vec
v,\omega)$
\begin{equation}\label{rho_t}
\rho(\vec v, t) = \int\rho(\vec v, \omega) \;e^{- \i \omega t} \;
\frac{d \omega}{2 \pi}.
\end{equation}
The function $\rho(\vec v,t)$ satisfies the following evolution
equation:
\begin{equation}\label{master_t}
\left(\frac{\partial}{\partial t} + \i \vec{kv}\right) \rho(\vec
v, t) =  \mathbf{\hat{S}} \rho(\vec v, t) + W(\vec v) \delta(t).
\end{equation}
This equation has only one solution that tends to zero at
$t\rightarrow\pm\infty$. Let us define the autocorrelation
function $\Psi(t)$ as
\begin{equation}\label{Phi(t)}
\Psi(t) = \int \rho(\vec v, t) \; d\vec v.
\end{equation}
This function is different from usually used function $\Phi(t)$
\cite{RS91} in the following way: $\Psi$ is an even function of
$t$, while $\Phi$ is zero at negative values of $t$. It is evident
from (\ref{rho_t}), (\ref{master_t}) and (\ref{Phi(t)}) that
$\Psi(t)$ is a real continuous even function connected with the
line shape by the Fourier transform
\begin{equation}\label{IPhi}
I(\omega) = \frac{1}{\pi}  \int \Psi(t) e^{i\omega t} dt .
\end{equation}
The asymptote of $I(\omega)$ at $\omega\rightarrow\infty$ is
determined by the discontinuities of the derivatives of $\Psi(t)$
at $t=0$. Using (\ref{master_t}) and taking into account
(\ref{killMaxwell}) and (\ref{killIntegral}) we find that (in the
absence of $\Gamma$) the lowest order of derivative of $\Psi(t)$
having a jump at $t=0$ is third. Thus the asymptote of $I(\omega)$
can be written as
\begin{equation}\label{Iasymp}
I(\omega) = \frac{1}{\pi \omega^4} \int d\vec v\; \vec k \vec v \;
\mathbf{\hat{S}} \; \vec k \vec v \;  W + O(1/\omega^6)
\quad\mbox{at}\; \omega\rightarrow\pm\infty.
\end{equation}
In order to obtain the value of the coefficient of the dominant term
of this asymptote we have to substitute the definition
(\ref{collisions}) of $\mathbf{\hat S}$  into Eq.(\ref{Iasymp}) with
some specific kernel $A(\vec v, \vec v')$. In the 1D approximation
we would have to substitute the reduced kernel (\ref{A1d}). It is
clear that the expression in 1D case is the same as in initial
model, the only difference is the order of integration. Thus the 1D
approximation always gives correct principal term of the asymptotic
expansion of the line shape at $\omega\rightarrow\infty$. It should
be noted that if $\Gamma$ is nonzero, the tails of the line have
universal Lorentzian form $\Gamma/\left(\pi \omega^2\right)$
regardless of any details of the collision integral.

The fact that the tails of the line profile are insensitive to the
transfer of non-equilibrium to the distribution over the
transverse velocities means that the deviation of the line shape
in 1D approximation manifests itself mostly close to the center of
the line. This tendency is evident, for instance, in
Fig.~\ref{fig:3d1d}, where the difference between 1D and 3D line
shapes is largest in the center. So in the rest of this paper we
will mainly analyze the behavior of intensity in the center of the
line $I_o=I(\omega=0)$.

\subsection{Small collision frequency}

The next limiting case we would like to consider is the case of
low collision frequency. In absence of the collision integral and
relaxation the line shape is Gaussian near the center:
\begin{equation}\label{I_small}
I^{(0)}(\omega)=\frac{1}{\sqrt{\pi} k
\vv_T}\exp{\left[-\frac{\omega^2}{(k \vv_T)^2}\right]} + O(\nu).
\end{equation}

If the collision frequency is small, the line shape $I(\omega)$
can be expanded in power series with respect to this parameter.
The first term of this series can be symbolically written as
\begin{equation}\label{I_small1}
I^{(1)}(\omega) = \int d\vec v \frac{1}{\i \omega - \i \vec k \vec
v} \mathbf{\hat{S}}\frac{1}{\i \omega - \i \vec k \vec v} W.
\end{equation}

Since $\mathbf{\hat{S}}$ is a linear integral operator, the
expression (\ref{I_small1}) contains integrations over both
initial and final velocities. If we perform the integration only
over the transverse velocities, we obtain the first term of this
expansion corresponding to 1D collision integral. Hence it is
evident that the first terms of the power expansion in
$\mathbf{\hat{S}}$ in 1D approximation and in initial model always
coincide.

\subsection{Small perturber mass $\beta\ll1$}

In this case the weak collision model is applicable. In this model
the collision integral is replaced by a Fokker-Planck differential
operator, and master equation takes the form:
\begin{equation}\label{weak}
\left( - \i \omega + \i \vec k \vec v\right)\rho = \nu
\left(\frac{\vv^2_T}{2} \Delta + \nabla\vec v\right) \rho.
\end{equation}
It is obvious that in this equation variables can be separated,
and thus the assumption of 1D approximation holds true. The
collision operator of the weak collisions model is really the
leading term of expansion of collision integral in power series in
small mass ratio $\beta$. In this section we are going to analyze
validity of 1D approximation in the next-to-leading order in
$\beta$.

This analysis should exactly account for collision frequency
$\nu$, because it is not supposed to be small, although it is
proportional to $\beta$. For instance, for the rigid spheres model
\begin{equation}\label{weak_nu}
\nu = \frac{8\sqrt{\pi}}{3}\;N_b\, \vv_{b T} \;a^2 \beta
\left(1+\beta\right)^2.
\end{equation}
Generally, the weak collisions model is applicable when $\beta$ is
small, but $N_b$ is large, so that $\nu$ is finite.

The term of the next-to-leading order in the collision operator is
a fourth order differential operator. For the rigid spheres it has
the form
\begin{multline}\label{L}
L= \beta \nu \left[ u^2 + \left(3+\frac{u^2}{5}\right) u \nabla +
\left(2+ \frac{3}{10}u^2 \right) \Delta +\right. \\ \left.+
\frac{3}{5} u_i u_k \nabla_i \nabla_k + \frac{4}{5} u \nabla
\Delta + \frac{1}{5} \Delta^2 \right],
\end{multline}
where $\vec u=\vec v/\vv_T$ and all the differentiations are with
respect to $\vec u$. In this expression, summation over repeating
indices is assumed. The next-to-leading term in the spectral line
shape (that is, exact in $\nu$ and of the first order in $\beta$)
can be calculated by considering this operator as a perturbation
and utilizing the time domain Green's function of equation
(\ref{weak}). But here we are interested only in the difference of
such terms in an arbitrary model and its 1D analog. This
difference can be symbolically written as
\begin{equation}\label{weak_perturb}
\delta \Psi^{(1)}(t) =\int\limits_0^t d t' \;\int\; d\vec v  \;
G(t-t') \left(L-L_{\mbox{\tiny 1D}}\right) G(t') W.
\end{equation}
Here $G(t)$ is the time domain Green's function of (\ref{weak}),
its velocity arguments are omitted, and $L_{\mbox{\tiny 1D}}$ is
one-dimensional fourth order differential operator representing
one-dimensional collision operator in the considered
approximation. Let us consider the expression (\ref{weak_perturb})
consequently. The expression $G(t') W$ (depending on velocity
$\vec v$ and time $t'$) is a product of Maxwellian distribution
over the transverse velocities and a non-equilibrium distribution
over the longitudinal velocity $\vv_z$. According to the
definition (\ref{A1d}), the integral over the transverse velocity
$\int d^2 \vec v_\perp L_{\mbox{\tiny 1D}} G(t') W$ is equal to
its 3D analog for any value of $\vv_z$:
\begin{equation}\label{zero_integral}
\int d^2 \vec v_\perp G(t-t') \left(L-L_{\mbox{\tiny 1D}}\right)\;
G(t')\; W = 0.
\end{equation}
The Green's function $G$ conserves the property of having zero
integral over $\vec v_\perp$ for all $\vv_z$, so the expression
(\ref{weak_perturb}) vanishes for all $t$. It means that 1D model
gives correct result for the correction of the first order in
$\beta$ to the weak collisions model.

\subsection{Heavy perturber $\beta \gg 1$}
\label{Lorenz}

The last limiting case we are going to consider is the Lorentz limit
$\beta \rightarrow \infty$ (see \cite{ufnR91}) of the rigid spheres
model. In this case the collision kernel and frequency are
\begin{equation} \label{ALorenz}
A(\vec v,\vec v')\rightarrow \frac{1}{2} N_{b}a^{2} \delta(\vec
v^2-\vec v^{'2}),\quad \nu(\vec v)=\pi N_b a^2 |\vec v|.
\end{equation}
In this case we can solve equation (\ref{master_omega}) for
$\omega=0$ and find the intensity in the center of the line:
\begin{equation}\label{Lorenz:I0}
I_o=\frac{2}{\pi^{3/2} k v_T} \;\; \frac{\mbox{arcctg}(C)}{1-C\,
\mbox{arcctg}(C)}\;,\quad\quad C=\frac{\pi N_b a^2}{k}.
\end{equation}
Constant $C$, being proportional to number density $N_b$ of the
buffer particles, differs from the collision frequency only by a
multiplier. Since our aim is only to compare 3D and 1D models, we
do not specify this factor in current section.

At large values of $C$ Dicke effect takes place, and the
asymptotic form of $I(0)$ is linear in collision frequency:
\begin{equation}\label{Lorenz:hydro}
\pi I_o = \frac{6\; C}{\sqrt{\pi} k v_T} + O(1/C)
\simeq 3.38 \frac{C}{k \vv_T}.
\end{equation}

The 1D collision kernel in the Lorentz limit is
\begin{equation}\label{Lorenz1d_A}
A(v_z |v'_z) = \frac{\pi}{2}N_ba^2 \left[ 1- \Theta\left(v_z -
v'_z\right)\left( 1- e^{-\left(v_z^2-v'^2_z \right)} \right)
\right].
\end{equation}
The out-scattering frequency in this approximation is
\begin{equation}\label{Lorenz1d_nu} \nu(v_z) = \pi N_ba^2 \left(v_z +
\frac{\sqrt{\pi}}{2}e^{v_z^2}\left(1-\mbox{Erf}(v_z)\right)
\right).
\end{equation}
The problem of finding  $I_o$ in this case can be reduced to
solving a second order ODE
\begin{equation}\label{L1D:differential}
\frac{d}{d\vv_z} \; e^{\vv_z^2} \; \left(\frac{k^2 \vv_z^2}{\nu} +
\nu \right)\frac{d f}{d \vv_z} = -\,2\, C\;  \vv_z\; e^{\vv_z^2}
f(\vv_z).
\end{equation}
with boundary conditions $f(\infty)=0$ and $f(0)=1$. Then the
intensity in resonance is given by
\begin{equation}\label{L1DI}
\pi I_o = - \left( \int_0^\infty \frac{\vv_z^2
f'(\vv_z)}{\nu(\vv_z)} \;d\vv_z\right)^{-1}.
\end{equation}
At large $C$ this expression takes the form
\begin{equation}\label{L1D:asymp}
\pi I_o = \left(\int W(\vv_z) \frac{k^2 \vv_z^2}{\nu_1(\vv_z)} \;
d\vv_z\right)^{-1} C + O(1/C)  \simeq 2.84 \frac{C}{k \vv_T}.
\end{equation}

\begin{figure}
\psfrag{abs}{\large$C$}\psfrag{ord}{\large$I(0)$}\includegraphics[width=0.7\columnwidth]{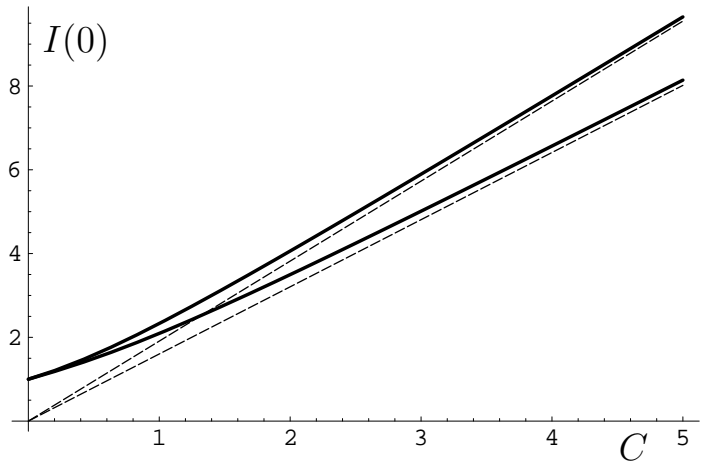}
\caption{The dependance of intensity $I_o$ in the center of the
line
 on collision frequency $C$ in the Lorentz limit. The upper
line corresponds to 3D Lorenz case, the lower --- to the 1D
approximation. The dashed lines represent linear asymptotes
(\ref{L1D:asymp}) and (\ref{Lorenz:hydro})}\label{fig:LorenzI0}
\end{figure}

The dependance of $I_o$ on $C$ obtained by numerical solution of
(\ref{L1D:differential}) together with dependance
(\ref{Lorenz:I0}) is presented on Fig.~\ref{fig:LorenzI0}. Both
functions increase monotonously with collision frequency,
approaching their linear asymptotes (\ref{Lorenz:hydro}) and
(\ref{L1D:asymp}).
Comparing (\ref{Lorenz:hydro}) and (\ref{L1D:asymp}) we find that
in the Lorentz limit the relative deviation of $I_o$ in the 1D
approximation is approximately $0.2$.

\section{Numerical results}\label{results}

The difference between the rigid spheres model and 1D
approximation is best characterized by  dependence of $I_o$ on the
collision frequency. Before calculating this dependence we have to
choose a value characterizing the collision frequency. In the
Lorentz limit we used $C$ to characterize it. In general case, it
is customary to use the diffusion frequency, defined as
$\nu_d={\vv_T}^2\left/{2 D}\right.$, where $D$ is the mutual
diffusion coefficient. The value of $D$ depends on the collision
kernel. Which kernel should we use to define $D$ and $\nu_d$? It
is clear that the 1D approximation should be compared with the
rigid spheres at the same values of physical parameters, such as
$N_b$ and $a$. Using its own diffusion coefficient for each model
would break this requirement. On the other hand, using the rigid
spheres diffusion coefficient for both models does not seem
consequent. Taking all that into account we decided to
characterize collision frequency by $\nu_{d}= {\vv_T}/{2 D'}$,
where $D'$ is the first order Chapman-Enskog diffusion
coefficient. The calculation of $D'$ only accounts for
longitudinal motion \cite{jcpL80}. Thus the 1D approximation leads
to correct value of $D'$ for any kernel. For the rigid spheres we
have
\begin{equation}\label{D'}
D'=\frac{ 3 \vv_T}{16 \sqrt{ \pi} a^2 N_b } \sqrt{\frac{1 + \beta
}{ \beta }}.
\end{equation}

In order to solve equations (\ref{master_omega}) with collision
kernel (\ref{Abilliard}) we used a method based on decomposition
of $\rho(\vec{v},\omega)$ into a linear combination of Burnett
functions \cite{praLindenfeld83,praCSDM02}. The problem is axially
symmetric, and can be reduced to two-dimensional by turning to
variables $x=v/v_{aT}$, $y=\cos(\vec{k},\vec{v})$.  The Burnett
functions have the form:
\begin{gather}
  \phi_{nl}(x,y) = N_{nl} x^l L_n^{l+1/2}(x^2)P_l(y), \\
  N_{nl} = \sqrt{\frac{\pi^{1/2}n!(2l+1)}{2\Gamma(n+l+3/2)}},\\
  L_n^{l+1/2}(x^2)=\sum_{m=0}^n\frac{(-1)^m\Gamma(n+l+3/2}{m!(n-m)!\Gamma(m+l+3/2}x^{2m},\\
  P_l(y)   =\frac{1}{2^l}\sum_{k=0}^{[l/2]}\frac{(-1)^k(2l-2k)!}{k!(l-k)!(l-2k)!}y^{l-2k},
\end{gather}
Here $N_{nl}$ is normalizing factor, $L_n^{l+1/2}(x^2)$ are
generalized Laguerre polynomials, $P_l(y)$ are Legendre
polynomials, $\Gamma(\dots)$ is Euler's Gamma function
\cite{Bateman}. The desired absorbtion intensity is given by
$n=0$, $l=0$ coefficient in the decomposition of
$\rho(\vec{v},\omega)$.

For numerical solution, we need to limit the decomposition by some
$l=l_{\max} $ and $n=n_{\max} $. The structure of emergent system
of linear algebraic equations allows to perform the calculation
considerably faster than for a generic linear system. The matrix
of the system turns out to be block tridiagonal if we group the
coefficients of decomposition of $\rho(\vec{v},\omega)$ so that
$l$ numerates blocks, and $n$ numerates the elements within each
block.  Then the blocks in the main diagonal are square symmetric
matrices, and the blocks in the neighboring diagonals are
two-diagonal matrices, the blocks above and below the main
diagonal are transposed with respect to each other. The vector in
the right hand side of the system has only one non-zero element,
$R_0=1$. Thus the system has the form
\begin{gather}
  À\vec{\rho}=\vec{R},\\
  \vec{R}=
  \begin{pmatrix}
   R\\
   0\\
   \vdots
  \end{pmatrix}
 A=\left(%
 \begin{array}{ccccc}
  \Lambda_0 & D_0        & 0         & 0  &  \cdots \\
  D^T_0     &  \Lambda_1 & D_1       & 0  &  \cdots \\
  0         & D^T_1      & \Lambda_2 & D_2&    \cdots\\
  \vdots    & \vdots     & \vdots    & \vdots  &\ddots\\
  \end{array}
\right).
\end{gather}

This system was solved by the special method for block tridiagonal
matrix \cite{G96}. As in the scalar marching method for a
tridiagonal matrix \cite{NR-92}, two sequences of coefficients are
calculated using recurrent formulas. The difference is, these
coefficients are not numbers but matrices and vectors. In our
case, only matrix sequence of coefficients needs to be calculated
because the vector coefficients are defined by the right hand side
and thus they are equal to zero:
\[
 \begin{array}{l}
  M_{l}=-\left(\Lambda_l+D_l M_{l+1} D^T_l\right)^{-1},\\
  M_{l_{max}}=-\Lambda^{-1}_{l_{max}}.
 \end{array}
\]
So, initial system is reduced to a system of dimensionality
$n_{max} +1$:
\[
 \left(\Lambda_0+D_0 M_1 D^T_0\right)\vec{\rho_0}=\vec{R_0}.
\]
This system was solved using the Givens rotation method \cite{G96}.

The one-dimensional problem (\ref{master1d}) was solved
numerically in a straightforward way by discretization of
velocity. The integrals appearing in (\ref{master1d}) were
approximated with sums by Simpson's formula with accuracy
$O(1/N^4)$ \cite{NR-92}, where $N$ is number of points. The
obtained system of linear equations was solved using Gaussian
method.

\begin{figure*}\centerline{
\subfigure[]{\psfrag{abs}{\large$\omega$}\psfrag{ord}{\large$I$}\includegraphics[width=0.33\textwidth,bb=50
45 410
302]{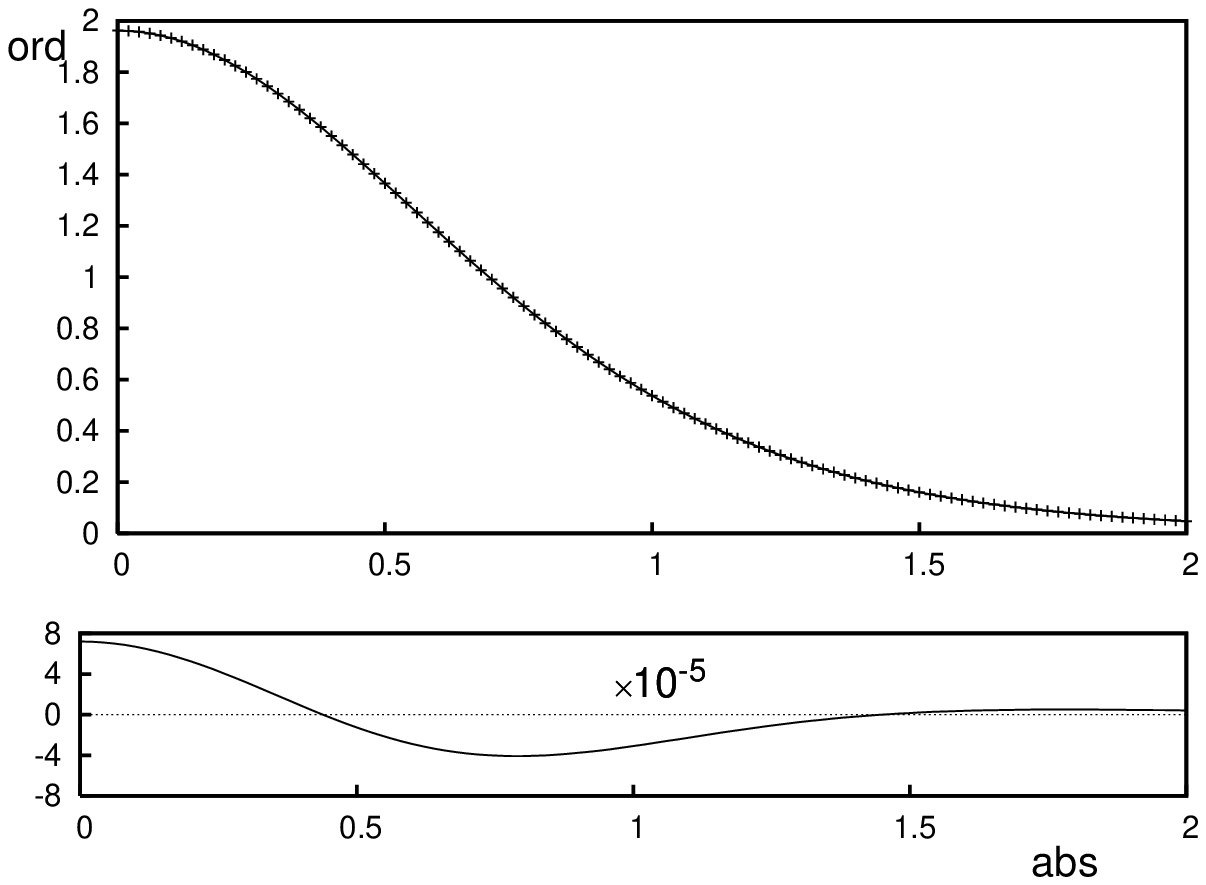}}\label{subfig:0203}\hfil%
\subfigure[]{\psfrag{abs}{\large$\omega$}\psfrag{ord}{\large$I$}\includegraphics[width=0.33\textwidth,bb=50
45 410
302]{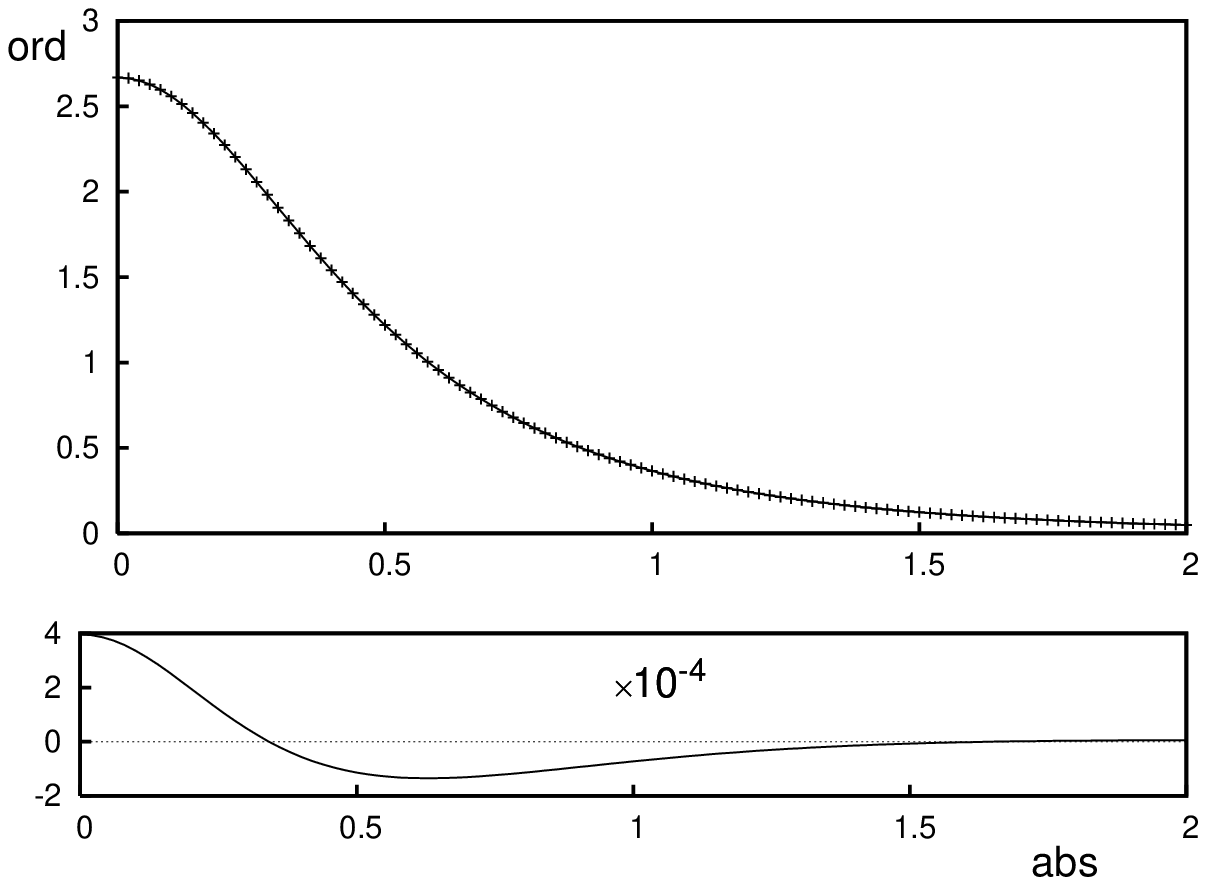}}\label{subfig:021}\hfil%
\subfigure[]{\psfrag{abs}{\large$\omega$}\psfrag{ord}{\large$I$}\includegraphics[width=0.33\textwidth,bb=50
45 410 302]{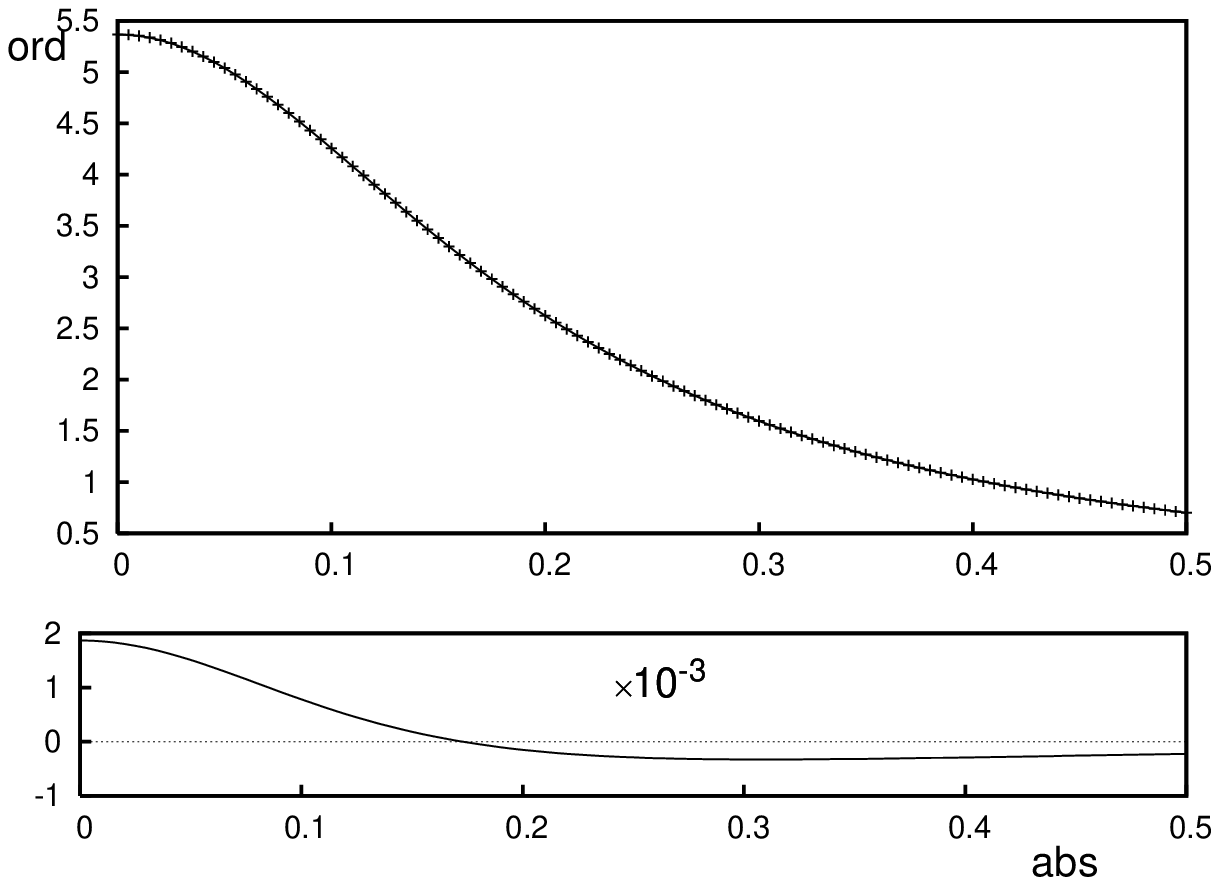}\label{subfig:023}}} \caption{The line
profiles in 3D and 1D rigid spheres model at $\beta=1/5$ and
$\nu_d=1/3 (a),1 (b),3 (c)$.}\label{fig:numerical1}
\end{figure*}

\begin{figure*}\centerline{
\subfigure[]
{\psfrag{abs}{\large$\omega$}\psfrag{ord}{\large$I$}\includegraphics[width=0.33\textwidth,bb=50
45 410
302]{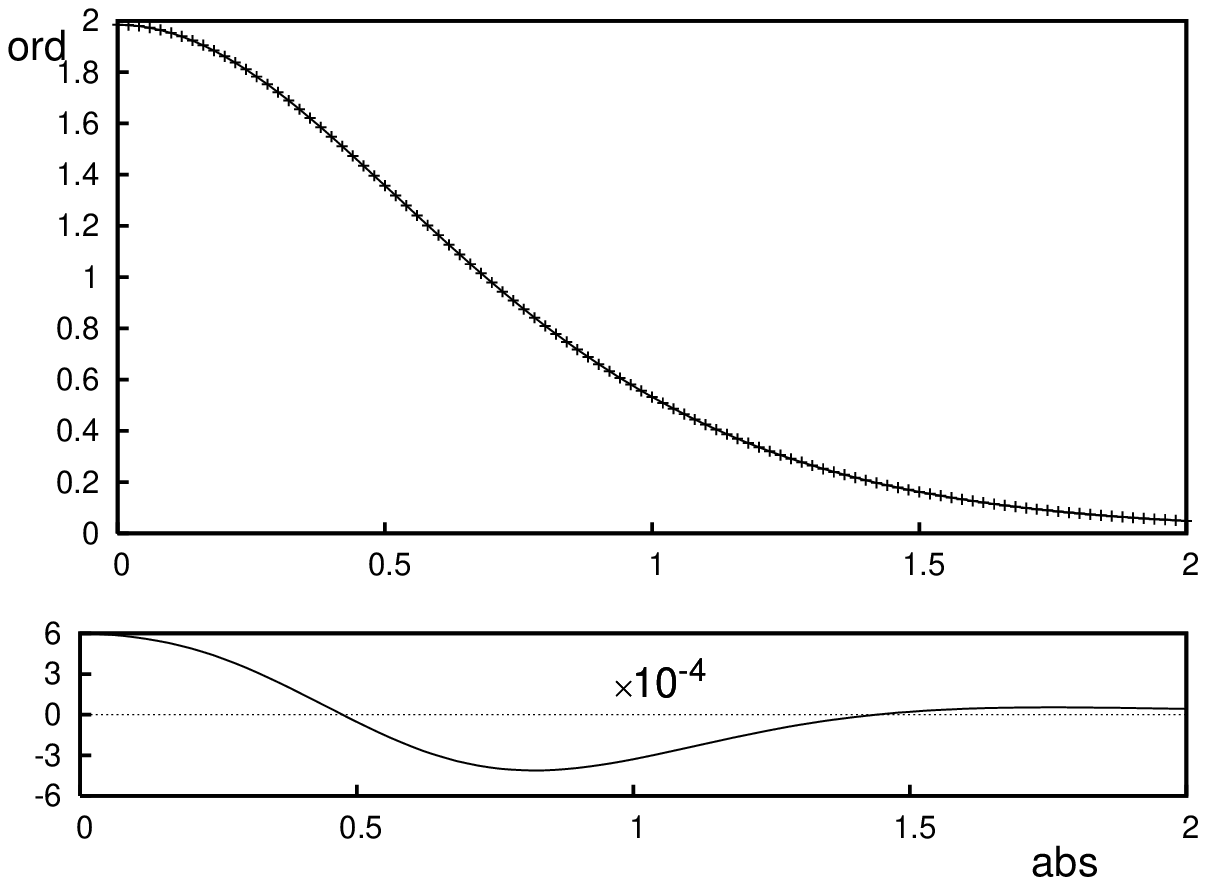}}\label{subfig:103}\hfil%
 \subfigure[]{\psfrag{abs}{\large$\omega$}\psfrag{ord}{\large$I$}\includegraphics[width=0.33\textwidth,bb=50 45 410
 302]{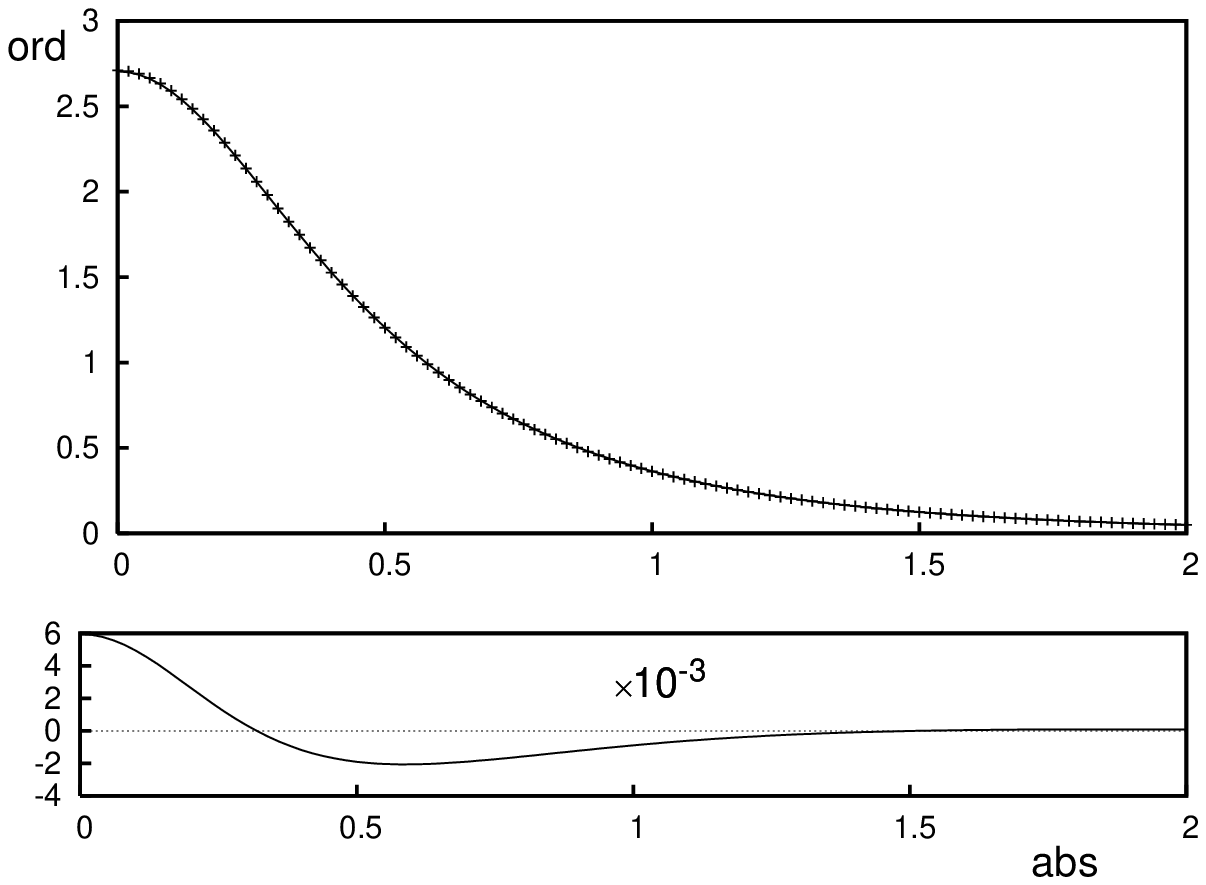}}\label{subfig:11}\hfil
 \subfigure[]{\psfrag{abs}{\large$\omega$}\psfrag{ord}{\large$I$}\includegraphics[width=0.33\textwidth,bb=50 45 410
 302]{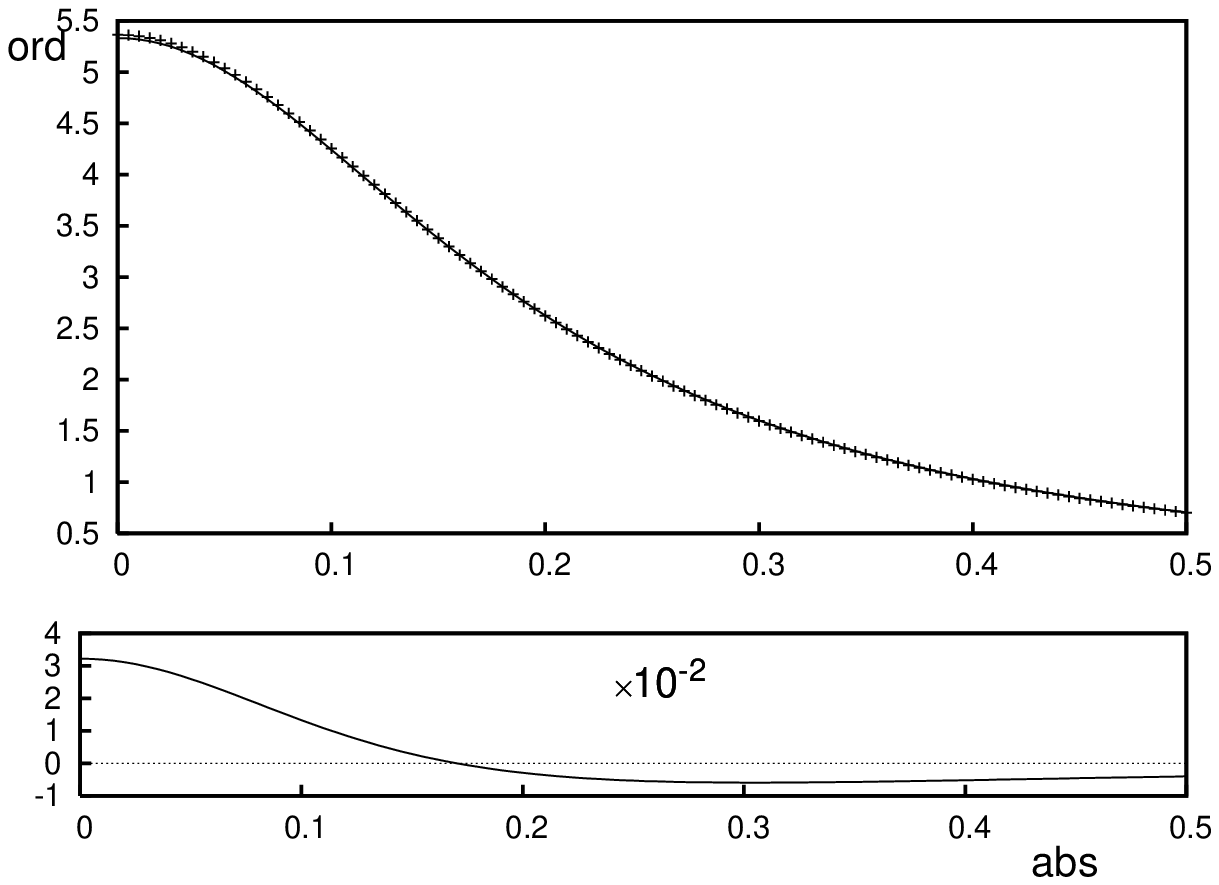}}\label{subfig:13}}
\caption{The same as in Fig.~\ref{fig:numerical1} at $\beta=1$.}
\label{fig:numerical2}
\end{figure*}

\begin{figure*}\centerline{
\subfigure[]{\psfrag{abs}{\large$\omega$}\psfrag{ord}{\large$I$}
\includegraphics[width=0.33\textwidth,bb=50 45 410
302]{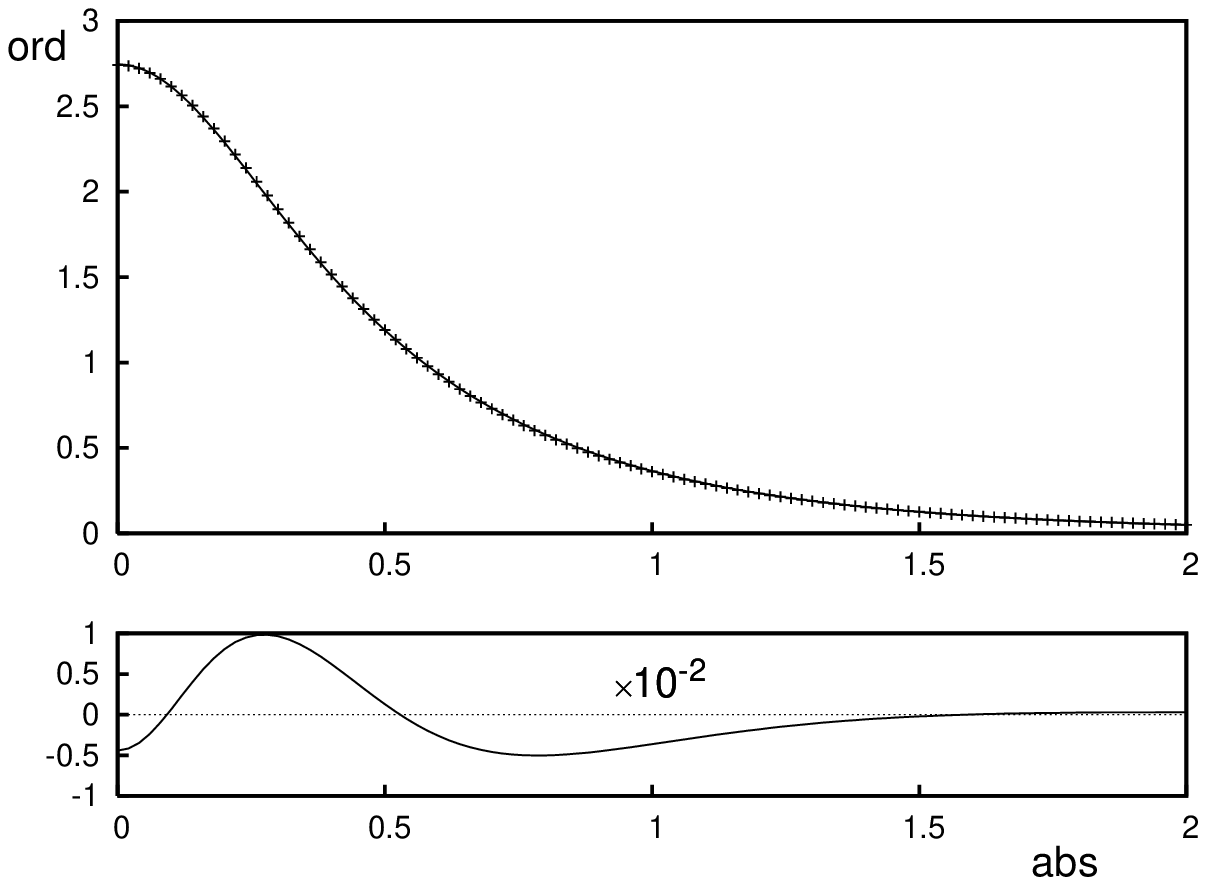}}\label{subfig:51}\hfil%
\subfigure[]{\psfrag{abs}{\large$\omega$}\psfrag{ord}{\large$I$}
\includegraphics[width=0.33\textwidth,bb=50 45 410
302]{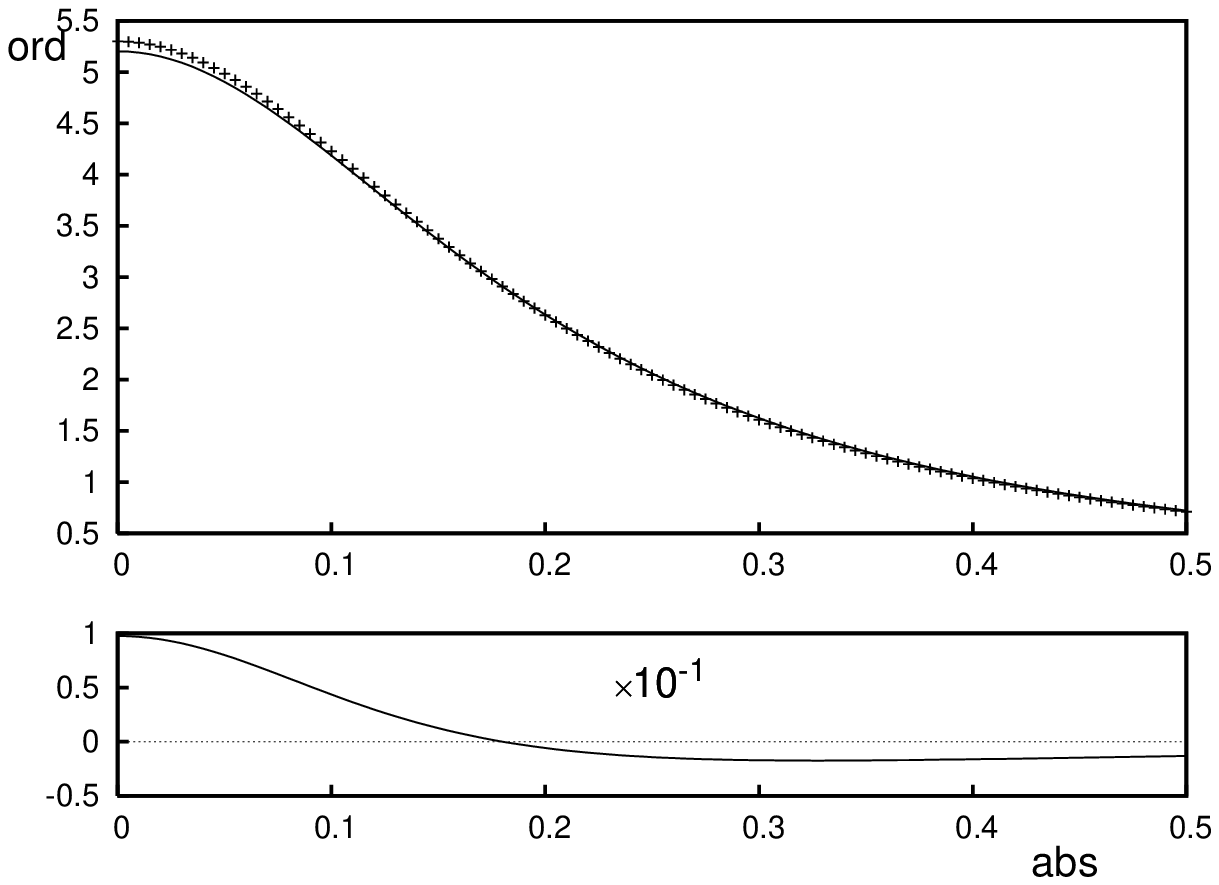}}\label{subfig:53}} \caption{The line profiles
in 3D and 1D rigid spheres model at $\beta=5$ and $\nu_d=1 (a),3
(b)$.} \label{fig:numerical3}
\end{figure*}

The results of 3D and 1D numerical calculations are presented on
Fig. \ref{fig:numerical1}, \ref{fig:numerical2},
\ref{fig:numerical3}. The lower panels show the difference between
1D and 3D calculations. The deviations are small, the
corresponding scaling coefficients are indicated. Fig.
\ref{fig:numerical1} corresponds to $\beta=1/5$, fig.
\ref{fig:numerical2} to $\beta=1$ and fig. \ref{fig:numerical3} to
$\beta= 5$. On the first two figures the three plots were
calculated at $\nu_d/k\vv_T = 1/3$, $1$, $3$, from left to right.
On the last figure the left plot corresponds to $\nu_d/k\vv_T=1$,
and the right one to $\nu_d/k\vv_T=3$. All profiles were
calculated at $\gamma=0.03$. The 3D plots were computed with
$l_{max}=n_{max}=48$.

The typical magnitude of deviation of 1D plots  on the three plots
of fig. \ref{fig:numerical1} is $10^{-5}$, $10^{-4}$, $10^{-3}$,
from left to right; on fig. \ref{fig:numerical2} it is $10^{-4}$,
$10^{-3}$, $10^{-2}$; and on fig. \ref{fig:numerical3} it is
$10^{-2}$ and $10^{-1}$. It can be seen that the difference
between line profiles in 3D and 1D case increases with $\beta$ and
with $\nu_d$. In agreement with theoretical expectations, the
deviation tends to zero at the wings of the line profiles.

Taking into account the behavior of deviation of 1D approximation
in the limiting cases, it seems natural to expect that the
relative deviation would grow monotonously with collision
frequency and with $\beta$, reaching maximum values of $\sim 0.2$.
However, this assumption proves to be wrong. Numerical
calculations in the rigid spheres model show that at $\beta < 100$
the relative deviation of 1D approximation is less than $0.1$. At
$\beta<30$ the deviation is less than $0.05$.

\begin{figure}
\psfrag{abs}{\large$\nu_d$}\psfrag{ord}{\large$I$}\includegraphics[width=0.8\columnwidth]{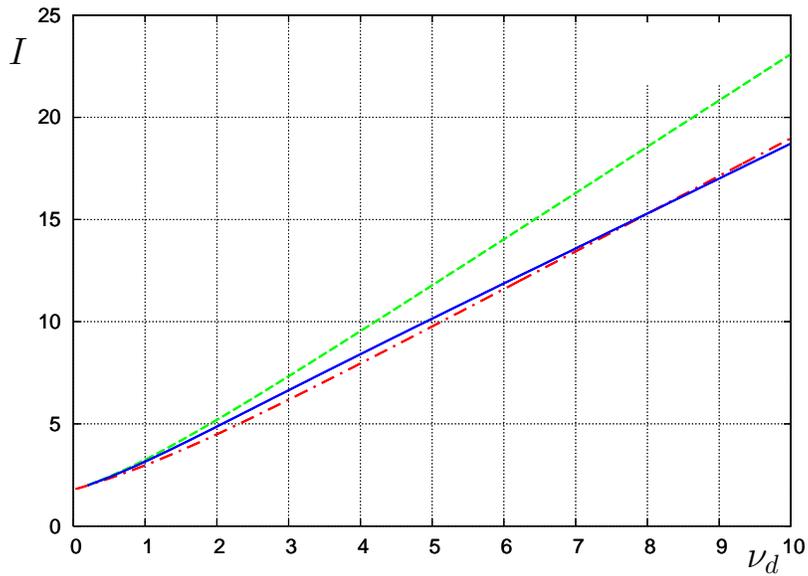}
\caption{The dependance of intensity in the center of the line
$I_o$ on collision frequency $\nu_d$ in the 3D and 1D rigid
spheres model for $\beta=100$ and in the Lorentz limit. The solid
line corresponds to 3D rigid spheres model at $\beta=100$, the
dot-dash one
--- to its 1D approximation. The dashed line corresponds to 3D
Lorentz limit. The 1D Lorentz curve is not plotted because it goes
very close to 1D with $\beta=100$.}\label{fig:beta100}
\end{figure}

Fig.~\ref{fig:beta100} presents $I_o(\nu_{d})$ obtained by
numerical calculation in the rigid spheres model and in its 1D
approximation for $\beta=100$ in comparison with the Lorentz case.
All the curves on this plot increase monotonously and approach
corresponding linear asymptotes as collision frequency tends to
infinity. The 3D $\beta=100$ curve is close to 3D Lorentz line at
small $\nu_d$. As the collision frequency increases, it goes down
and crosses the 1D line, representing both Lorentz and $\beta =
100$ case.

The behavior of the curves on this plot can be described as
follows. The Lorentz model is the limiting case of the rigid
spheres model at $\beta\rightarrow\infty$. The dependence of $I_o$
on $\nu_{d}$ in the rigid spheres model approaches that dependence
in the Lorentz limit as $\beta$ goes to infinity. However, this
approaching is not uniform in $\nu_{d}$: at any fixed $\beta$, no
matter how large, at sufficiently large $\nu_{d}$ the dependence
deflects from Lorentzian asymptote downwards and approaches some
other linear asymptote. This other (final) asymptote goes lower
than (\ref{L1D:asymp}), so the 3D line crosses the one
corresponding to 1D approximation. It can be said that due to this
intersection of the two plots the error of 1D approximation at
reasonable values of $\beta$ is much less than in Lorentz limit.

As it is known, in case of full phase memory in the limit of large
collision frequency the line shape takes the form
\begin{equation}\label{hydro:I}
I(\omega) =\frac{1}{\pi} \;\; \frac{ D k^2}{D^2 k^4 + \omega^2}
\;\; +\;\; O(D).
\end{equation}
In case of large $\beta$ this formula gives
\begin{equation}\label{general:Iasymp}
\pi I_o = \frac{9 \pi}{16} \frac{\nu_d}{(k\vv_T)^2} + O(1/\nu_d)
\simeq 1.77 \frac{\nu_d}{(k \vv_T)^2}.
\end{equation}
This formula  describes the asymptotic behavior of the rigid
spheres model at large but finite $\beta$. It perfectly agrees
with numerical results. However, formula (\ref{general:Iasymp})
does not describe the Lorentz limit itself, its asymptote is given
by (\ref{Lorenz:hydro}). The derivation of (\ref{hydro:I}) implies
that the only distribution that is turned to zero by the the
collision operator is equilibrium distribution. The Lorentzian
collision operator is degenerate in this sense, because it turns
to zero all distributions depending only on absolute value of
velocity. This consideration shows that in case of large but
finite $\beta$ function $I_o(\nu_{d})$ should behave as follows:
while $\nu_{d}$ is not too large, it is close to
(\ref{Lorenz:I0}). At large values of $\nu_{d}$ the function
$I_o(\nu_{d})$ must approach the asymptote (\ref{general:Iasymp}).
This is exactly what numerical calculations show. The remaining
question is, at what values of $\nu_{d}$ does the function
$I_o(\nu_{d})$ switch from (\ref{Lorenz:I0}) to
(\ref{general:Iasymp}).

Let us consider the collision operator of the rigid spheres model.
As it is shown in \cite{jAM81}, all its eigenvalues are positive,
except the one corresponding to equilibrium distribution, which is
zero. It is clear that in case of large $\beta$ the minimal
positive eigenvalue has the order of magnitude of $\beta^{-1}$. It
means that the coefficient in $O(1/\nu_{d})$ term in
(\ref{hydro:I}) is of the order $\beta$. Thus the first term in
(\ref{hydro:I}) is much greater than the second when $\nu_{d}^2 /
(k \vv_T)^2 \gg \beta$. So the transition happens (and the 1D plot
intersects with rigid spheres plot) at $\nu_{d} \simeq
\sqrt{\beta} \, k \vv_T$. This result is in good agreement with
numerical calculations.

As it was mentioned, the most significant difference between 3D
and 1D line profile is observed in case which is effectively
Lorentzian: $1\ll\nu_d/ (k \vv_T)\ll\sqrt{\beta}$. The reason of
such big deviation is that at large $\beta$ the 3D rigid spheres
model and 1D approximation behave differently: in 3D model, the
kinetic energy of an active molecule almost does not change in
collisions. In contrast to that, in 1D approximation there is no
such conservation. This "energy persistence" property of the 3D
rigid spheres collision kernel can be interpreted in terms of the
generalization of the Keilson-Storer model recently proposed in
\cite{epjBonamy_TranThiNgoc_Joubert_Robert}. This model introduces
two velocity persistence parameters, $\gamma_m$ and $\gamma_o$.
They are responsible for the persistence of the modulus of
velocity and its orientation, correspondingly. The separation of
these two parameters is incident in the rigid spheres model.
Indeed, in case of heavy perturber gas, the typical change of the
speed of an active molecule in a single collision is of the order
$\vv_T \beta^{-1}$, while its orientation changes totally. Thus,
in case of large $\beta$ the modulus persistence parameter
$\gamma_m$ is close to unity, and the difference has the order of
magnitude
\begin{equation}\label{gamma_m}
1-\gamma_m \approx \beta^{-1}.
\end{equation}
This simple estimate agrees remarkably well with the figures
obtained in \cite{epjBonamy_TranThiNgoc_Joubert_Robert} by
simulation: for $H_2$ in nitrogen ($\beta \approx 14$) and in
argon ($\beta \approx 20$) the  estimate (\ref{gamma_m}) gives
$\gamma_m = 0.93$ and $\gamma_m = 0.95$, while the figures
presented in \cite{epjBonamy_TranThiNgoc_Joubert_Robert} are
$0.92$ and $0.96$ correspondingly.

\section{Conclusion}\label{conclusion}

We have analyzed several limiting cases for arbitrary collision
kernel and found the following. The term $1/\omega^4$ in the
asymptote of the tails of the line shape, which is principal in
absence of dephasing, is given correctly by the 1D approximation.
Thus the transfer of disequilibrium distribution to the transverse
components of velocity manifests itself mostly close to the center
of the line, and the error of 1D approximation can be
characterized by the deviation of absorption in the center of the
line. We demonstrated that in case of small collision frequency
the first correction in this parameter is also given correctly by
1D approximation.

In the limit of small perturber to radiator mass ratio $\beta$ it is
known that the line shape in 1D approximation coincides with that in
the initial 3D model. We have shown that the line shape given by 1D
approximation remains correct also in the next order in $\beta$. In
the opposite limiting case of large $\beta$ any realistic 3D
collision integral conserves the kinetic energy of active molecules.
The 1D collision integral cannot reproduce such property. We have
shown that due to this difference, the most significant deviation of
1D line shapes is observed in case of heavy perturber molecules. The
deviation vanishes at small collision frequency, and reaches its
maximal value in the hydrodynamical limit $\nu \gg k\vv_T$. We found
that in the rigid spheres model for infinite $\beta$ the relative
error of 1D approximation reaches the value of $\sim 20\%$.

The case of intermediate mass ratio was analyzed numerically in
the rigid spheres model. The inaccuracy of 1D approximation was
found considerably smaller than expected value of $\sim 20\%$.
This happens due to intersection of the plots describing the
collision frequency dependance of intensity in the center of the
line in 1D and 3D models. For moderate values of $\beta$, that is,
$\beta\lesssim\beta_0 \sim 30$, the relative error of 1D
approximation does not exceed $0.05$. The error increases
monotonously with $\beta$. At $\beta < \beta_0$ the error takes
its maximal value in the hydrodynamical limit. At larger $\beta$
the maximal error occurs at $\nu\sim \sqrt{\beta} k\vv_T$.

Thus the one-dimension model could be applied for light
perturbers, at low pressure or in the problems where 5\% is
sufficient accuracy. For arbitrary buffer particles or precision
calculation the three dimension billiard ball approximation
becomes preferable.

\section*{Acknowledgements}
The authors are grateful to A.D.~May, S.G.~Rautian and
A.M.~Shalagin for helpful discussions. The work is partially
supported by the Program of the Physical Sciences Department of
Russian Academy of Sciences.


\newpage
\end{document}